\documentclass{moriond}

\usepackage{amssymb}
\usepackage{amsmath}
\usepackage{tikz}
\usetikzlibrary{shapes,arrows,positioning,automata,backgrounds,calc,er,patterns}
\usepackage[compat=1.1.0]{tikz-feynman}
\usepackage{slashed}

\def\be{\begin{equation}}
\def\ee{\end{equation}}
\def\bea{\begin{eqnarray}}
\def\eea{\end{eqnarray}}

\newcommand{\Photo}{}

\begin{document}
\vspace*{4cm}
\title{THERMAL EFFECTS IN $\nu$DM PRODUCTION}

\author{S.~ROSAURO-ALCARAZ~\footnote{In collaboration with A.~Abada, G.~Arcadi, M.~Lucente and G.~Piazza.}}

\address{P\^ole Th\'eorie, Laboratoire de Physique des 2 Infinis Irène Joliot Curie (UMR 9012), \\
CNRS/IN2P3,
15 Rue Georges Clemenceau, 91400 Orsay, France.}

\maketitle\abstracts{We study the possibility to produce a keV neutrino dark matter candidate through the two-body decays of heavy neutrinos present in TeV scale neutrino mass generation mechanism. Given that the dark matter production happens at the heavy neutrino scale, namely around the electroweak scale, we address thermal effects and study how these modify the dark matter production rates through freeze-in.}

\section{Introduction}
Thanks to numerous gravitational probes at different scales, we know there is dark matter (DM) in our Universe. Its relic density, as measured by Planck~\cite{Planck:2018vyg}, represents about $25\%$ of the energy budget of the Universe, $\Omega_{DM}h^2=0.1198\pm0.0012$. However, all our efforts to find conclusive evidence for DM interaction with Standard Model (SM) particles in our direct or indirect detection experiments have so far failed. It is thus interesting to consider DM species which interact very feebly with the SM. Given that neutrinos are the most weakly interacting particles present in the SM and that we need to explain the origin of their mass with new physics, in the following we will consider the possibility that the new states generating light neutrino masses also play a role in the DM production in the Early Universe.

\section{Origin of neutrino masses}
Among the many different scenarios which could account for the smallness of neutrino masses, one particularly appealing possibility is that of low-scale seesaws, in which neutrino masses are protected by an approximate lepton number symmetry, while the heavy states can have masses at the $\mathcal{O}\left(100\right)$~GeV in reach of colliders. One such example is the so-called inverse seesaw~\cite{Gonzalez-Garcia:1991brm,Malinsky:2005bi} (ISS), in which we add a number of right-handed (RH) neutrinos, $N_R$, and singlet fermions, $S$, to the SM particle content, whose lagrangian is enlarged with
\begin{equation}
    \mathcal{L}\supset -\bar{L}_L Y_{\nu}\tilde{\Phi}N_R-\bar{S}MN_R-\frac{1}{2}\bar{S}\mu S^c + h. c.,
    \label{eq:lagrangian}
\end{equation}
where $L_L$ is the SM lepton doublet and $\tilde{\Phi}\equiv i\sigma_2\Phi$ is the Higgs doublet and $Y_{\nu}$ the neutrino Yukawa coupling. Given that lepton number is assumed to be an approximate symmetry of the theory, and the term proportional to $\mu$ is the only source of its breaking, we assume $\mu$ to be much smaller than $M$ and the electroweak (EW) scale. This is indeed technically natural as the symmetry protects it from receiving important radiative corrections, and thus $\mu$ stays small at all orders in perturbation theory. Upon spontaneous symmetry breaking (SSB), one finds that light neutrino masses are proportional to
\begin{equation}
    m_{\nu}\sim v^2 Y_{\nu}^T M^{-1}\mu M^{-1}Y_{\nu},
\end{equation}
where $v$ is the Higgs vev. Note, however, that the mixing between the active and the heavy neutrinos is given by the ratio $\theta\sim v Y_{\nu}M^{-1}$, which can be large. It was found in Ref.~[4]~\cite{Abada:2014vea} that the minimal number of RH neutrinos and singlet fermions needed to explain oscillation data was two, while a DM candidate of mass $\mathcal{O}(\mu)$ naturally appears when including yet another singlet fermion~\cite{Abada:2014zra}. This case was labelled as $(2,3)$-ISS, pointing out the number of RH neutrinos and singlet fermions, respectively. Upon full diagonalization of the neutrino mass matrix in the latter case, we find the linear relation between the interaction states from Eq.~(\ref{eq:lagrangian}) and the mass eigenstates
\begin{equation}
    \begin{pmatrix} 
    \nu_L \\ N_R^c \\ S 
    \end{pmatrix}=\mathcal{U}P_L
    \begin{pmatrix}
        n_{light}\\
        n_{DM}\\
        n_{heavy}
    \end{pmatrix},
\end{equation}
where we have neglected explicit flavour indices and ordered the mass eigenstates, $n_i$, according to the scale of the masses, $m_{\nu}$, $\mu$ and $M$, respectively. The mixing between light neutrinos and DM would be given by the mixing matrix element $\mathcal{U}_{\alpha 4}$, with $\alpha=e,\mu,\tau$.

\section{DM production through freeze-in}
It was found in Ref.~[5]~\cite{Abada:2014zra} that in the $(2,3)$-ISS induced neutrino mass spectrum, between the light neutrinos participating in oscillations and the two pseudo-Dirac pairs with masses $\mathcal{O}(M)$, there is a sterile neutrino with mass $\mathcal{O}(\mu)$, whose interactions are automatically suppressed by powers of neutrino mixing, $\mathcal{U}_{\alpha 4}$, thus making it a good warm DM candidate when $\mu\sim$~keV. Given its feebly interactions, it can be produced through two-body decays of the heavier neutrinos in the early Universe, and it would not produce a detectable signal in our direct detection experiments. Instead, the optimal way to look for this DM candidate is through radiative decay into lighter neutrinos and a photon, emitting a monochromatic X-ray signal~\cite{Pal:1981rm}. Indeed, the non-observation of the monochromatic signal in astrophysical observations sets severe constraints on the size of the mixing~\cite{Foster:2021ngm,Boyarsky:2007ge,Roach:2019ctw}, down to $|\mathcal{U}_{\alpha 4}| \lesssim 10^{-6}$.

When thermal effects can be neglected, the relic abundance of a DM species produced via freeze-in through two-body decays of a heavier mother particles can be estimated as
\begin{equation}
    \Omega_{DM}h^2\sim 3\times 10^{24}\frac{m_{DM}\Gamma(N\rightarrow \nu_{DM} +X)}{m_N^2},
\end{equation}
where $m_N$ is the heavy neutrino mass, $X$ is any particle resulting from the heavy neutrino, $N$, decay and $m_{DM}$ is the DM mass. One can check that for a typical $m_N\sim 150$~GeV and $m_{DM}\sim 10$~keV, the decay rate needs to be of the order of $\Gamma(N\rightarrow \nu_{DM} +X)\sim 10^{-16}$~GeV, which can in principle be achieved for example when the heavy neutrino decay into DM and a Higgs boson. Other production channels have been studied before in the literature\cite{Merle:2014xpa,DeRomeri:2020wng,Lucente:2021har}. In the following, we will however study how thermal effects~\cite{le_bellac:1996} can drastically change the production of neutrino DM in the Early Universe, very much in line with the study of Ref.~[14]~\cite{Lello:2016rvl}.

\begin{figure}
\centering
\begin{tikzpicture}
  \begin{feynman}
    \vertex (a) {$n_i$};
    \vertex [right=1cm of a] (b);
    \vertex [right=3cm of b] (c);
    \vertex [right=1cm of c] (f2) {$n_j$};

    \diagram[horizontal=a to b]{
      (a) -- [plain] (b),
      (b) -- [plain, edge label=$n_k$, momentum'=$q$] (c),
      (b) -- [scalar, edge label=$H$, half left, momentum'=$p-q$] (c),
      (c) -- [plain] (f2),
    };
  \end{feynman}
\end{tikzpicture}
\begin{tikzpicture}
    \begin{feynman}
    \vertex (a) {$n_i$};
    \vertex [right=1cm of a] (b);
    \vertex [right=3cm of b] (c);
    \vertex [right=1cm of c] (f2) {$n_j$};

    \diagram[horizontal=a to b]{
      (a) -- [plain] (b),
      (b) -- [plain, edge label=$n_k$, momentum'=$q$] (c),
      (b) -- [boson, edge label=$Z$, half left, momentum'=$p-q$] (c),
      (c) -- [plain] (f2),
    };
  \end{feynman}
\end{tikzpicture}
\caption{Feynman diagrams contributing to the neutrino self-energy. Note that the $W$ boson also contributes with a charged fermion running in the loop. Its structure can be readily extracted from the $Z$ one, with the appropriate changes of masses and couplings.}
\label{fig:self-energies}
\end{figure}
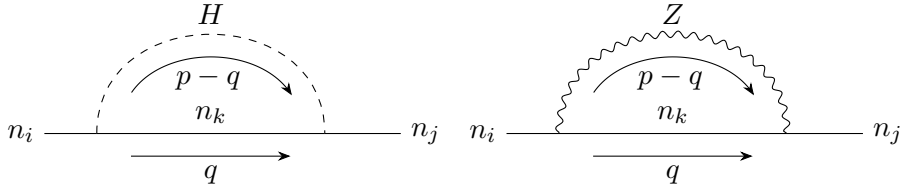
In order to completely capture the thermal effects affecting neutrino propagation at finite temperature, it is necessary to consider the neutrino self-energy corrections at finite temperature. To that purpose we use the \textit{real-time formalism}~\cite{le_bellac:1996} to obtain the self-energy corrections in the mass basis, which can in general be written as
\begin{equation}
    \Sigma_{ij}(p_0,|\vec{p}|,T)=\sum_{k}C_{ik}C_{kj}\sigma(p_o,|\vec{p}|,T,m_k),
    \label{eq:self-energy}
\end{equation}
where the sum runs over mass eigenstates with masses $m_k$ and $C_{ij}\equiv \sum_{\alpha} U^{\dagger}_{i\alpha}U_{\alpha j}$. The function $\sigma$ contains all the dependence on the temperature and the neutrino momenta, $(p_0,\vec{p})$. Its particular form is not important for the following discussion. This correction will translate into a modification of the dispersion relation for neutrinos, through
\begin{equation}
    \sum_{j}\left[\slashed{p}-\mathcal{M}+\Sigma(T)\right]_{ij}n_j=0,
\end{equation}
where $\mathcal{M}$ is the diagonal mass matrix and we have only explicitly written the dependence of $\Sigma$ on the temperature. In practice, one needs to find the new propagating states in the medium, $\mathcal{N}_i(T)$, which differ for different helicities~\cite{Lello:2016rvl}, thus obtaining an effective mixing angle as
\begin{equation}
    n_i=\sum_{p}\mathcal{A}_{ip}(T)\mathcal{N}_p(T)\rightarrow \mathcal{V}(T)\equiv\mathcal{U}\mathcal{A}(T).
\end{equation}

For the sake of illustration, it is useful to consider the ``toy case'' in which we have just one SM neutrino and the light sterile DM, for which the Higgs contribution can be neglected and the neutrino masses can be neglected when evaluating the function $\sigma$ from Eq.~(\ref{eq:self-energy}). In this case we can analytically find the effective mixing matrix in the medium, $\mathcal{V}(T)$. In particular, we can define an effective mixing angle as
\begin{equation}
    \theta^h_{eff}(T)\equiv \frac{\mathcal{U}_{\alpha 4}}{\sqrt{\left((1+\left(\frac{\text{Re}\Omega^h(T)}{m_{DM}^2}\right)\right)^2+\left(\frac{\text{Im}\Omega^h(T)}{m_{DM}^2}\right)^2}},
\end{equation}
where $\Omega^h(T)\sim(p_0-h|\vec{p}|)\sigma$ and $h=\pm1$ denotes the helicity of the neutrino. It is thus clear that, in the relativistic limit, right-handed helicity neutrinos ($h=+1$) will not receive large self-energy corrections, and thus the mixing angle in the medium is approximately the one in vacuum, while for left-handed helicity neutrinos ($h=-1$) these corrections will be important and thus the mixing angle will in general be suppressed. 

Finally, the production rate for the DM can be written as
\begin{equation}
    \Gamma_{DM}^h(T,|\vec{p}|)\sim 2\left(\theta^h_{eff}(T)\right)^2\text{Im}\left[\Omega^h(T)\right],
    \label{eq:effective_Mix}
\end{equation}
from which one realizes that there will be a competition between the suppression of the effective mixing angle with the temperature and the generally larger damping rate at larger temperatures. Additionally, the production rate for RH helicity neutrinos, even if proportional to the mixing angle in vacuum, will also be suppressed by the chirality flipping, $\Omega^h(T)\sim (p_0-|\vec{p}|)\sigma\sim m_{DM}^2\sigma/2|\vec{p}|$~\cite{Lello:2016rvl}.

\begin{figure}
    \centering
    \includegraphics[width=0.75\textwidth]{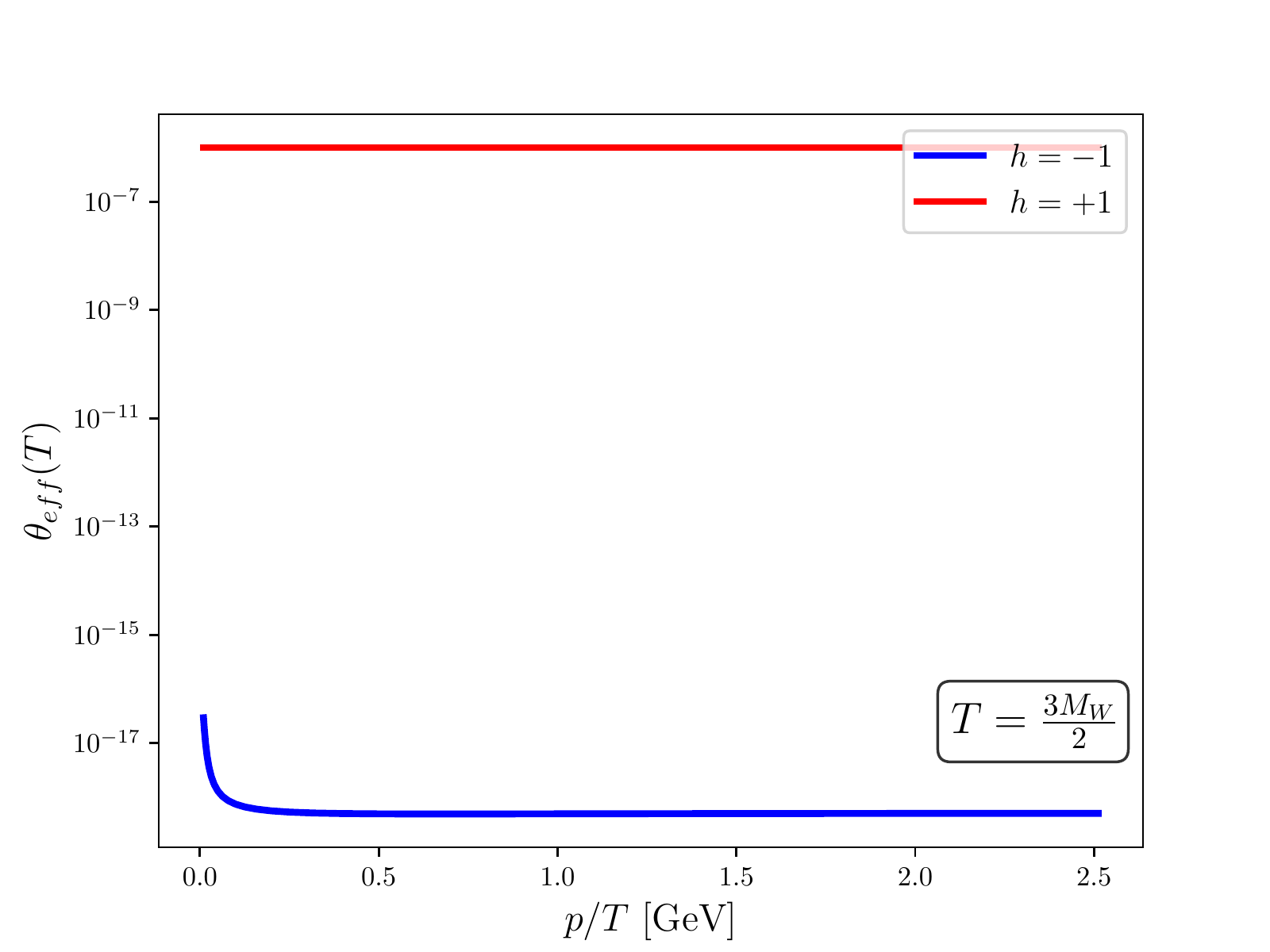}
    \caption{Effective mixing angle for RH helicity (red) and LH helicity (blue) neutrinos as a function of the ration between the momenta $p\equiv |\vec{p}|$ and the temperature. The mixing angle in vacuum is set to $\left|\mathcal{U}_{\alpha 4}\right|\sim 10^{-6}$.}
    \label{fig:effective_mixing}
\end{figure}

In Fig.~\ref{fig:effective_mixing} we show the difference in the effective mixing angle, using the ``toy'' two-neutrino scenario, assuming they are Dirac particles. One can observe how for the left-handed (LH) helicity neutrinos, $h=-1$, the effective mixing angle is reduced by orders of magnitude (blue line), while for RH helicity neutrinos its value is very close to the vacuum one, which has been set in this example to $\mathcal{U}_{\alpha 4}\sim 10^{-6}$.
However, from Eq.~(\ref{eq:effective_Mix}) we also note that, if $\text{Re}\Omega^h(T)\sim -m_{DM}^2$, it is possible to have a resonance which might enlarge the production rate, although it is always partially dumped by $\text{Im}\Omega^h(T)$. 

\begin{figure}
    \centering
    \includegraphics[width=0.48\textwidth]{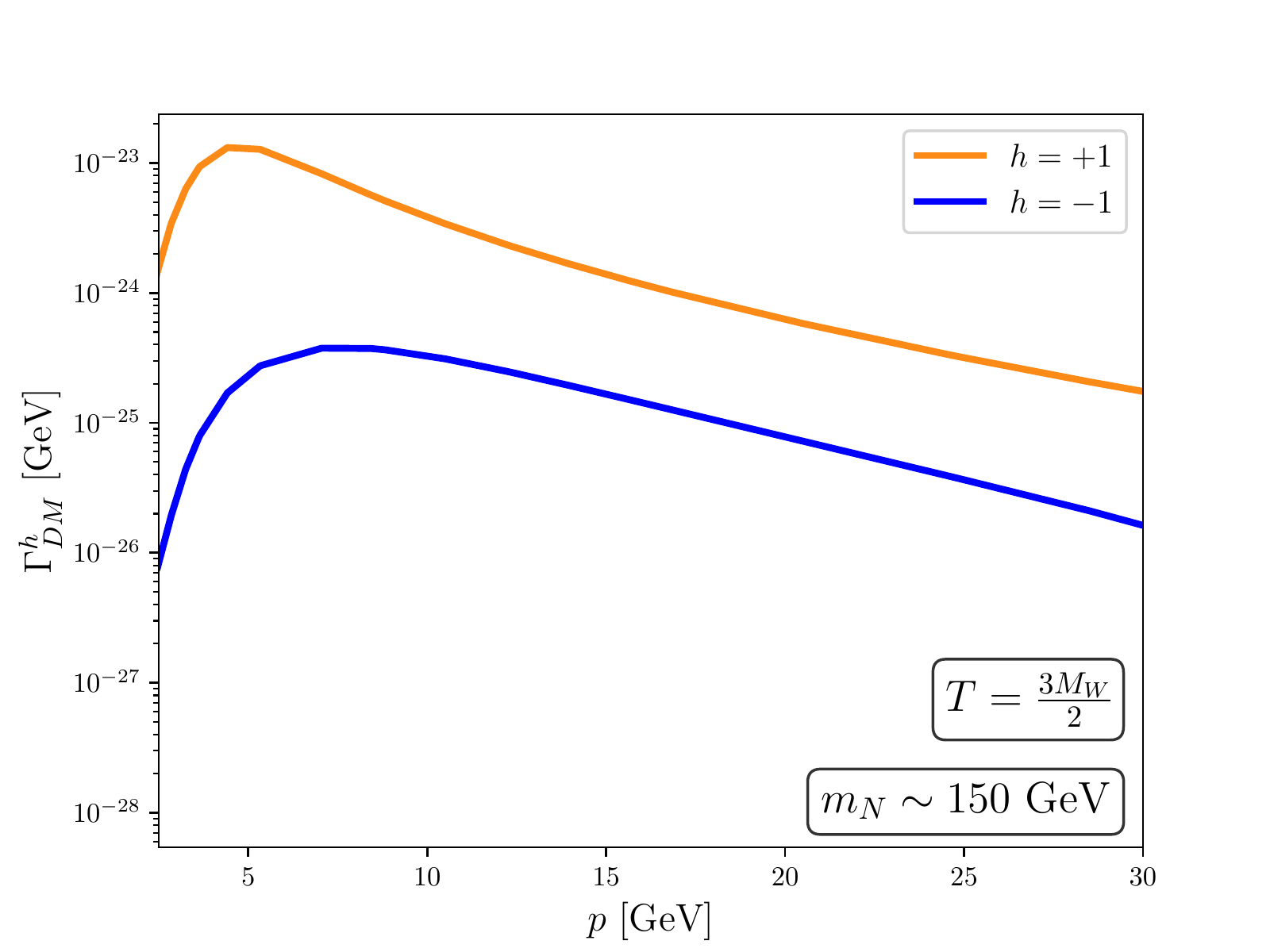}
    \includegraphics[width=0.48\textwidth]{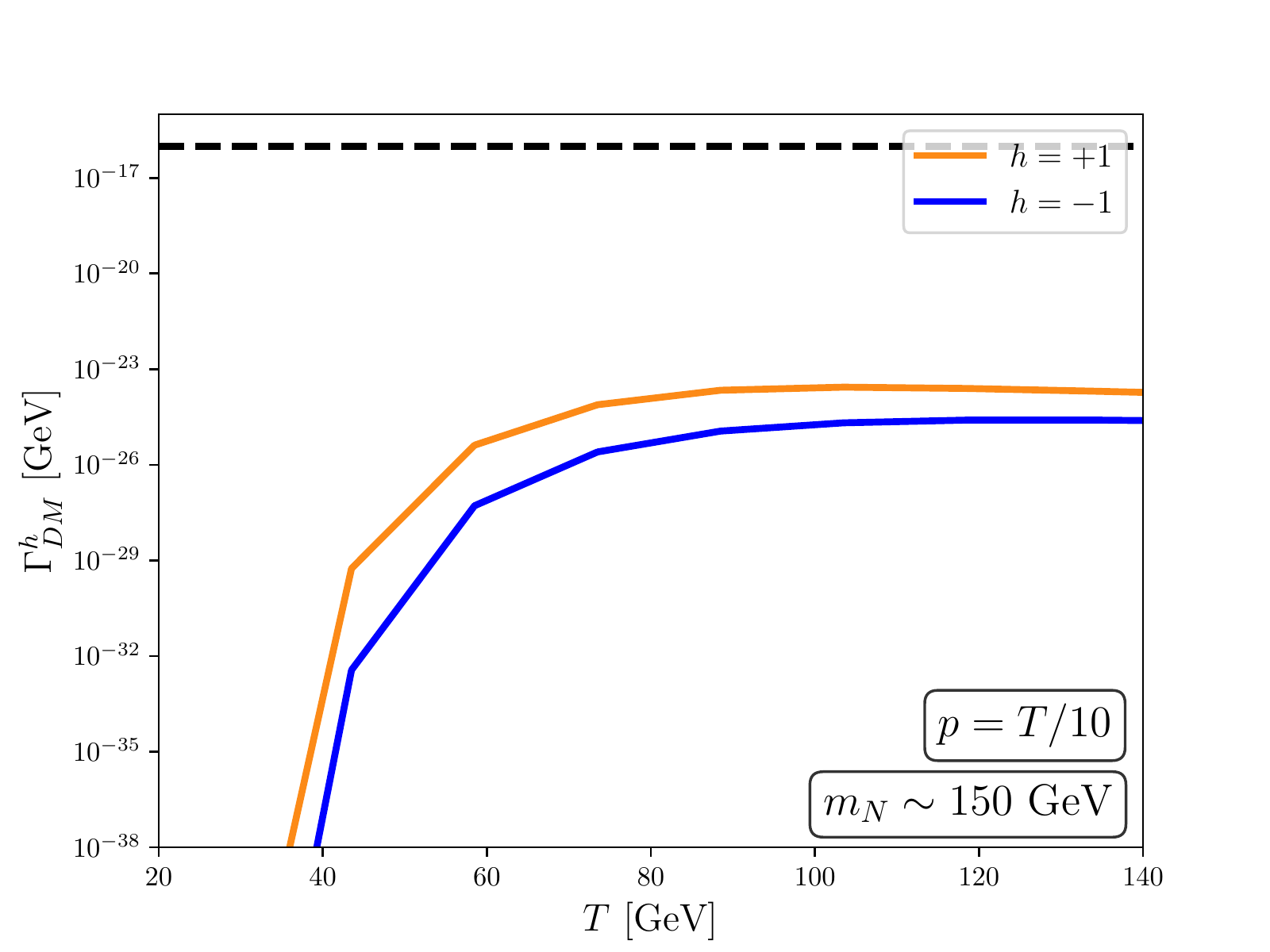}
    \caption{Total production rates for a keV-neutrino DM candidate through the decays of a heavier neutrinos into a boson and the DM candidate , $N\rightarrow \nu_{DM}+B$, and subsequent decays of said bosons, $B\rightarrow \nu_{DM}+f$, where $f$ is either a light neutrino or a charged lepton. The blue (orange) line corresponds to LH (RH) helicity neutrinos. The left panel shows the dependence on the momenta for a fixed temperature while the right one the dependence on the temperature. The scale of the heavy neutrinos has been chosen close to the EW scale, $m_N\sim\mathcal{O}(150)$~GeV.}
    \label{fig:production_rates}
\end{figure}

The inclusion of the heavy neutrino contributions, together with the Higgs self-energy contribution, help enlarge the production rates and thus the final DM abundance~\footnote{ Including the heavy neutrinos, it is no longer possible to write the effective mixing angle in the medium analytically}. This is indeed what we find, in particular, for the LH helicity neutrinos, for which the production rates are enlarged as compared with those in Ref.~[14]~\cite{Lello:2016rvl} and which we show in Fig.~\ref{fig:production_rates}. We show in blue and orange the full production rates for DM below the EW scale for RH and LH helicity neutrinos, respectively. As can be seen, even if there is a difference between both helicities, it is not as large as the one found in Ref.~[14]~\cite{Lello:2016rvl}. This is due both to the inclusion of the heavy neutrinos and the Higgs contributions to the production, as well as to the fact that in the ISS scenario taken here as an example, neutrinos are Majorana particles and thus the production is slightly altered. The right panel shows additionally the ``naive'' production rate one would need to explain the whole DM abundance as a dashed black line. In the context of freeze-in, we should solve the Boltzmann equation
\begin{equation}
    \frac{d n^h_{DM}}{dt}=\Gamma^h_{DM}(T,|\vec{p}|) n_{eq}(T)
\end{equation}
to find the total abundance and really compare with the ``naive'' expectation, but it seems clear that it is very difficult to explain the whole DM abundance just by adding the heavy neutrinos are their decays into DM, due to the strong thermal effects that hinder the production.

\section{Conclusions}
We have studied the production of a keV-neutrino DM candidate arising from the neutrino mass mechanisms, for example, in the context of low-scale seesaw. In particular, we have scrutinized the unavoidable thermal effects that appear just from the interactions of the DM with the rest of particles in the plasma, mediated through neutrino mixing. This same neutrino mixing controls the production rate, and thus there will be a competition between the stronger interactions suppressing the effective mixing in the medium and enlarging the production rate. We find that, while the addition of the heavy neutrinos which can decay into the DM and a Higgs boson improve the final DM abundance, it does not seem to be enough to really explain the whole Dm relic density. On the other hand, taking into account the full mixing matrix with all possible CP-violating phases might improve the situation by taking into account possible interference between diagrams. On the other hand, the production through some non-minimal coupling not relying on mixing and not completely related to the neutrino mass generation is still possible.

\section*{Acknowledgments}

The authors acknowledge support through the European Union's Horizon 2020 research and innovation programme under the Marie Sklodowska-Curie grant agreements No 860881-HIDDeN and No 101086085-ASYMMETRY.

\end{document}